\pdfoutput=1
\documentclass{aastex61}

\received{November 28, 2016}
\accepted{January 30, 2017}
\submitjournal{ApJ}
\usepackage{natbib}
\shorttitle{Suppression of Hydrogen Emission in an X-class White-light Solar Flare}
\shortauthors{Proch\'{a}zka et al.}

\begin{document}
\title{Suppression of Hydrogen Emission in an X-class White-light Solar Flare}
\correspondingauthor{Ond\v{r}ej Proch\'{a}zka}
\email{oprochazka01@qub.ac.uk}
\author{Ond\v{r}ej Proch\'{a}zka}
\affil{Astrophysics Research Centre, Queen's University Belfast, Northern Ireland, UK}
\author{Ryan O. Milligan}
\affil{Astrophysics Research Centre, Queen's University Belfast, Northern Ireland, UK}
\affil{NASA Goddard Space Flight Centre, Greenbelt, MD 20771, USA}
\affil{Department of Physics, Catholic University of America, 620 Michigan Avenue, Northeast, Washington, DC 20064, USA}
\author{Joel C. Allred}
\affil{NASA Goddard Space Flight Centre, Greenbelt, MD 20771, USA}
\author{Adam F. Kowalski}
\affil{Department of Astrophysical and Planetary Sciences, University of Colorado Boulder, 2000 Colorado Ave, Boulder, CO 80305, USA}
\affil{National Solar Observatory, University of Colorado Boulder, 3665 Discovery Drive, Boulder, CO 80303, USA
}

\author{Pavel Kotr\v{c}}
\affil{Astronomical Institute, The Czech Academy of Sciences, 25165 Ond\v{r}ejov, Czech Republic}
\author{Mihalis Mathioudakis}
\affil{Astrophysics Research Centre, Queen's University Belfast, Northern Ireland, UK}

\begin{abstract}
We present unique NUV observations of a well-observed X-class flare from NOAA 12087 obtained at Ond\v{r}ejov Observatory. The flare shows a strong white-light continuum but no detectable emission in the higher Balmer and Lyman lines. RHESSI and Fermi observations indicate an extremely hard X-ray spectrum and $\gamma$-ray emission. We use the RADYN radiative hydrodynamic code to perform two type of simulations. One where an energy of 3 $\cdot$ 10$^{11}$ erg cm$^{-2}$ s$^{-1}$ is deposited by an electron beam with a spectral index of $\approx$ 3 and a second where the same energy is applied directly to the photosphere.  The combination of observations and simulations allow us to conclude that the white-light emission and the suppression or complete lack of hydrogen emission lines is best explained by a model where the dominant energy deposition layer is located in the lower layers of the solar atmosphere rather than the chromosphere. 
\end{abstract}
\section{Introduction}
One of the main characteristics of solar white-light flares (WLF) is strong emission in the higher order hydrogen Balmer lines and the near UV continuum (\citealt{Fletcher:2015aa}, \citealt{Donati-Falchi85}). WLF can be classified as type I or type II according to their spectral properties, which correspond with a prevailing emission mechanism \citep{Machado:1986aa}. Type I WLF have strong, broad Balmer lines emission and are thought to originate in the chromosphere at a temperature of about $10^4$ K. In contrast type II WLF show weaker Balmer line emission, no evidence of a Balmer edge at 364.6 nm and are thought to originate deeper in the photosphere with a density higher than $10^{15}$ cm$^{-3}$ and a strong H$^-$ contribution. 
 \cite{Ding:1999aa} highlighted that the models of type II WLF require a significant temperature increase around the upper photosphere and temperature minimum region. They speculate that the heating mechanism should keep the chromosphere relatively undisturbed and that the flare energy must be deposited in the lower layers of the solar atmosphere (also see \citealt{Li:1997aa}). In a related study, \cite{Metcalf:1990aa} used two neutral magnesium lines (457.1 nm and 517.3 nm) to quantify the changes in the atmospheric structure during five flares (all with electron spectral index smaller than 4) and concluded that in three of these events the only heating and ionization mechanism was chromospheric backwarming by Balmer and Paschen continuum radiation. They also reported that in two of the events, the Mg I 517.3 nm line core emission did not correlate with X-rays and the cause of the enhancement in the Balmer and Paschen continua was unclear.
 
In a WLF observation close to the solar limb (N05E64), \cite{Boyer:1985aa} measured that the flare excess signal related to the quiet solar signal around the Balmer jump reached up to 19\%. Strong emission in Ca II H\&K and H$\epsilon$ lines was also detected. They concluded that Paschen or H$^-$ continuum radiation were not likely to be responsible for the emission and  instead proposed the presence of a slightly warmer layer ($\approx$ 150 K) in the photosphere. \cite{Hudson:2010aa} compared the three observations of WLF Balmer continua presented by \cite{Hiei:1982aa}, \cite{Neidig:1983aa} and \cite{Machado:1974aa}. Their spectral signatures varied from the clear presence of a Balmer jump to one that was shifted towards longer wavelengths and to its complete absence.

A comprehensive study of flares on M dwarf stars was recently carried out by \cite{Kowal2013} who acquired a large number of spectra over the spectral range 340 to 920 nm. The wide spectral coverage allowed the simultaneous study of the Balmer lines and the Balmer and Paschen continua.  One of the common features of the flares was a Balmer continuum in emission. The non-detection of the Balmer jump in the high resolution spectra (R $\sim$ 40,000), was attributed to the presence of a pseudo-continuum caused by the  blending of the higher-order Balmer lines. The stellar flare observations (400 - 480 nm) also indicate a hot blackbody emission with characteristic temperatures of around 10,000 K. One of the most surprising results was the detection of strong absorption hydrogen Balmer and Ca II H\&K lines in the impulsive phase of an M-dwarf megaflare, in strong resemblance to the spectrum of hot stars (i.e. Vega). This indicates that the heating takes place in the deep layers of the atmosphere.  

The first attempts to model the response of the solar atmosphere to an injected electron beams date back to 1980s. \cite{Zharkova:1993aa} presented a detailed analysis of the response of the Balmer and Lyman lines as well as continua on electron beams with different initial fluxes and spectral indices. They point out that both the Lyman and Balmer line wings appear to rise in models with higher electron beam fluxes ($10^{11}$ erg cm$^{-2}$ s$^{-1}$ compared to $10^9$ erg cm$^{-2}$ s$^{-1}$) and lower spectral indices. Line core intensities tend to decrease.

In the present paper we analyze multi-wavelength observations of a rare X-class flare. Our observations include the hydrogen Balmer and Lyman lines and continua. A comparison with  radiative hydrodynamic simulations is also presented.

\section{Observations \& Data Reduction}
On 11 June 2014 a solar flare classified by the \textit{Geostationary Operational Environmental Satellites} (GOES) as X1.0 peaking at 9:06 UT was observed in active region NOAA 12087 near the eastern limb (S18E57). The event was observed  at Ond\v{r}ejov Observatory, Czech Republic, the \textit{Solar Dynamics Observatory} (SDO), \textit{Reuven Ramaty High-Energy Solar Spectroscopic Imager} (RHESSI, \citealt{Lin:2002aa}) and the \textit{Fermi Gamma-ray Space Telescope} (\citealt{Meegan:2009aa}) satellites.
This flare was the fourth strong event in this active region detected over the previous 24 hours. The preceding day two other events appeared (X2.2 peaking at 11:42 UT and X1.5 peaking at 12:52 UT) followed by M3.0 event on 11 June peaking at 8:09 UT.\\
\subsection{Optical spectroscopy}
\label{optical}
We obtained spectra in the 350 - 485 nm wavelength range at Ond\v{r}ejov Observatory using the \textit{Horizontal Sonnen Forschungs Anlage 2} (HSFA 2, \citealt{Kotrc:2009aa}) telescope with a Jensch-type coelostat. The instrument consists of an image selector with a set of circular diaphragms delimiting the active region under investigation, a 1D spectrometer HR4000 by Ocean Optics, Inc. with a spectral resolution of $\sim$0.03 nm per pixel, and an H$\alpha$ context imaging system. The spectrograph has been optimized for observations of strong solar flares at sub-second resolution. After going through the selected diaphragm, the light is focused onto an optical fibre which feeds the spectrometer. The outer rim of the diaphragm reflects light into a camera allowing us to receive a context image of the target that emits the detected spectrum. The spectrometer includes a built-in 14-bit CCD with very low noise characteristics. The mean value of 50 dark frames with exposure time of 30 ms is 668.18 $\pm$ 0.16 counts with a standard deviation of 9.89 $\pm$ 0.13 over the whole wavelength range. During the recording we used a function of the spectrometer's control software \textit{Electric Dark Correction}, which allows us to subtract the read-out offset automatically. A set of dark frames was recorded every 20 - 60 minutes with the same exposure time as the observations. A quiet Sun spectrum at approximately the same distance from the disc centre was also recorded for comparison. The quiet Sun data can provide a reference spectrum that allows us to estimate the excess emission during the flare. The spectral data can be displayed as wavelength versus counts plots or be used to construct lightcurves in 11 pre-defined channels. A detailed description of this spectrograph has been published by \cite{Kotrc:2016aa}.

\begin{figure}[p]
\centering
\includegraphics[width=0.85 \textwidth]{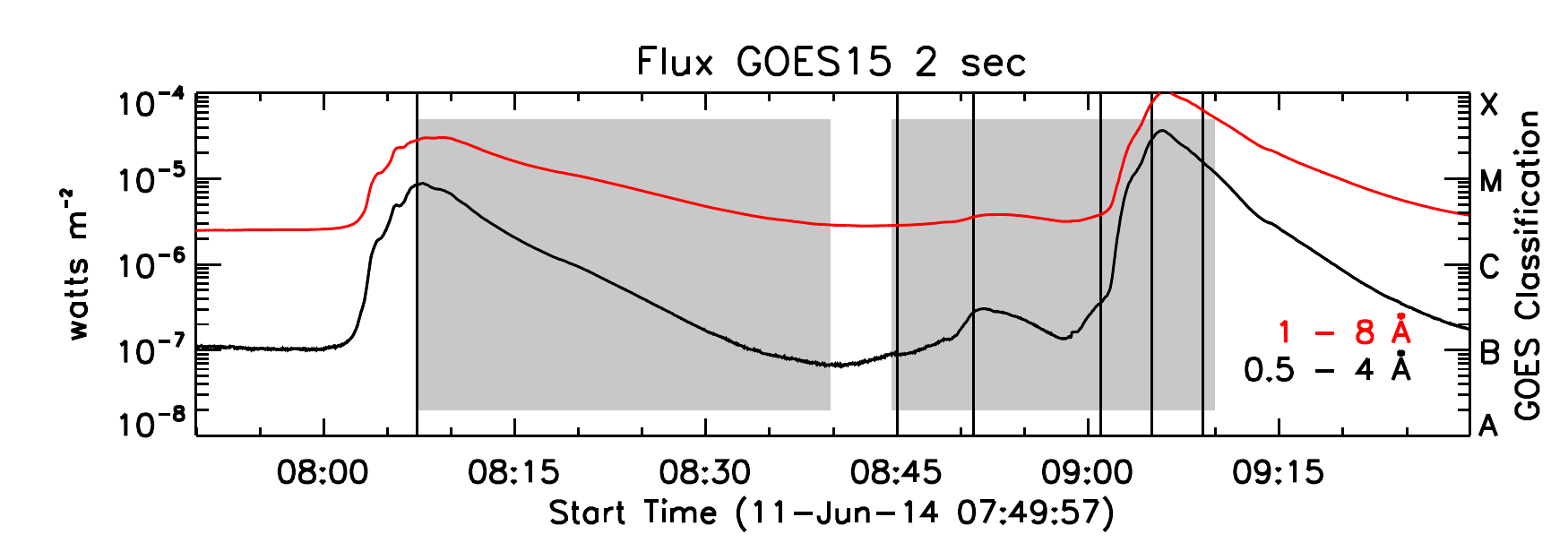}
\includegraphics[width=0.85 \textwidth]{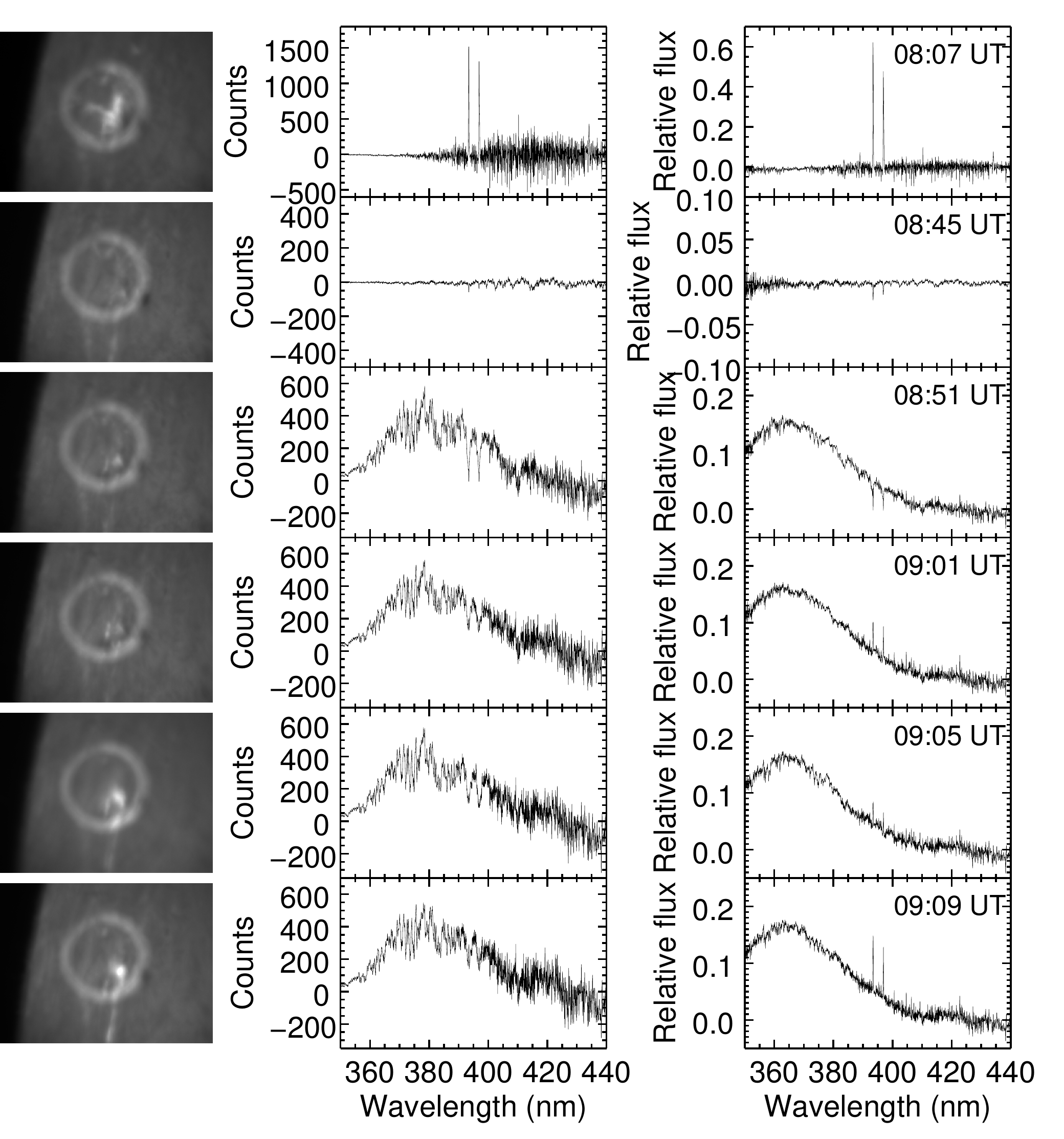}
\caption{Upper panel: GOES X-ray lightcurves for the flares under investigation. The grey background marks the time intervals covered by spectroscopy in the visible. Lower panel: Spectra at selected times (also marked in the upper panel by vertical black lines) along with a context H$\alpha$ images. The overlaid bright circle is a reflection from the delimiting aperture with a diameter of 10 mm, which corresponds to 57$''$ in the focal plane. The middle column shows the net flare spectra (i.e. a quiet Sun spectrum has been subtracted).  The right column shows the spectra in the middle column normalized to quiet Sun.} 
\label{fig2}
\end{figure}

The observation of AR 12087 commenced on 11 June at 8:07 UT with an integration time of 40 ms and a cadence of 0.09 seconds (Figure \ref{fig2}). The observations were interrupted briefly at 8:39 UT for calibration purposes 
and recommenced at 8:44 UT with an integration time 30 ms. The data acquisition ended at 9:09 UT. We used a diaphragm with diameter of 10 mm (57 arcsec). Both datasets were supported by a context imaging in H$\alpha$ line with a cadence of 0.2 and 1 s, respectively.

\subsection{SDO/HMI white-light emission}
\begin{figure}[h]
\centering
\vspace{-0.5cm}
\includegraphics[width=0.5 \textwidth]{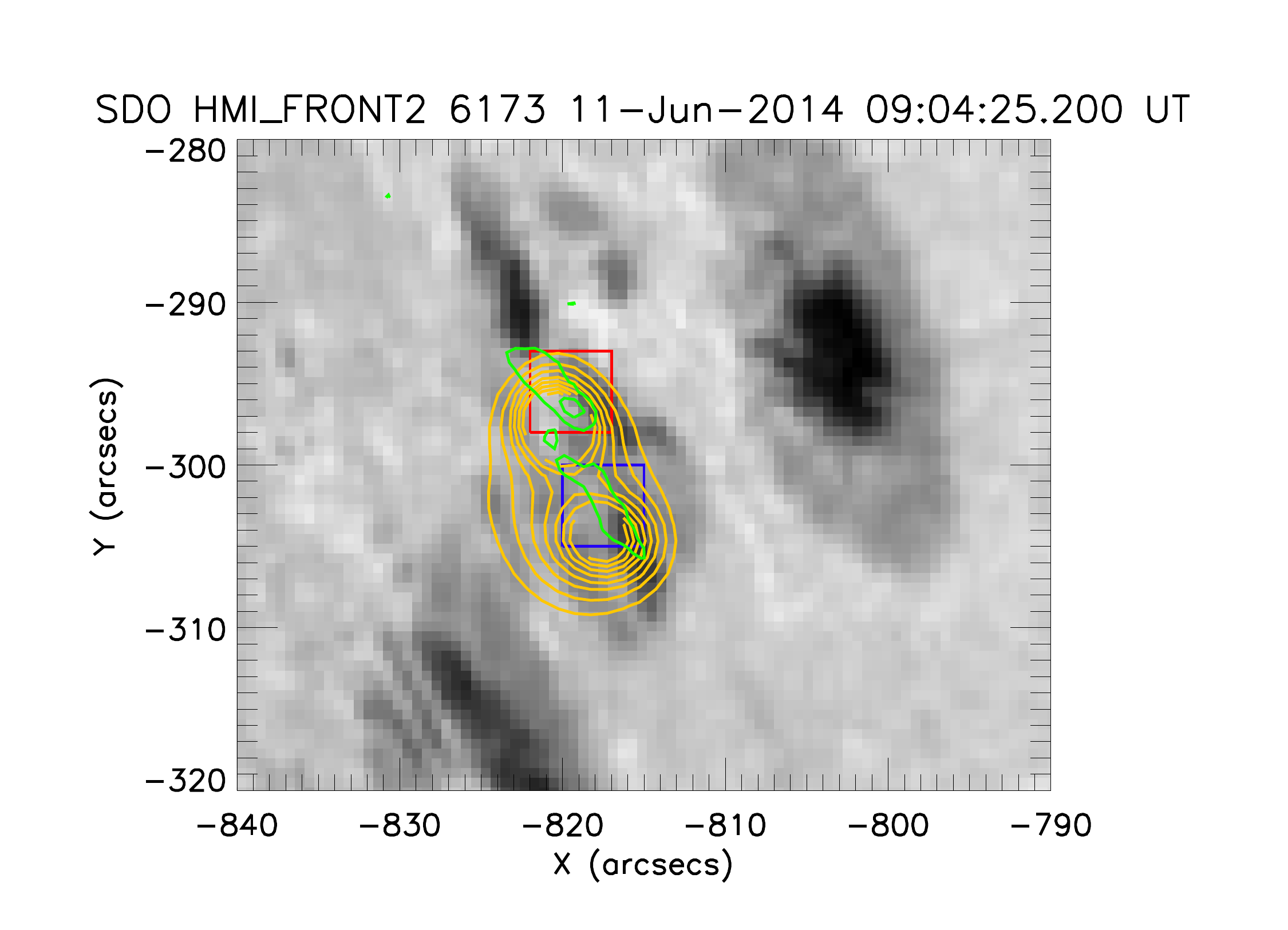}
\includegraphics[width=0.45 \textwidth]{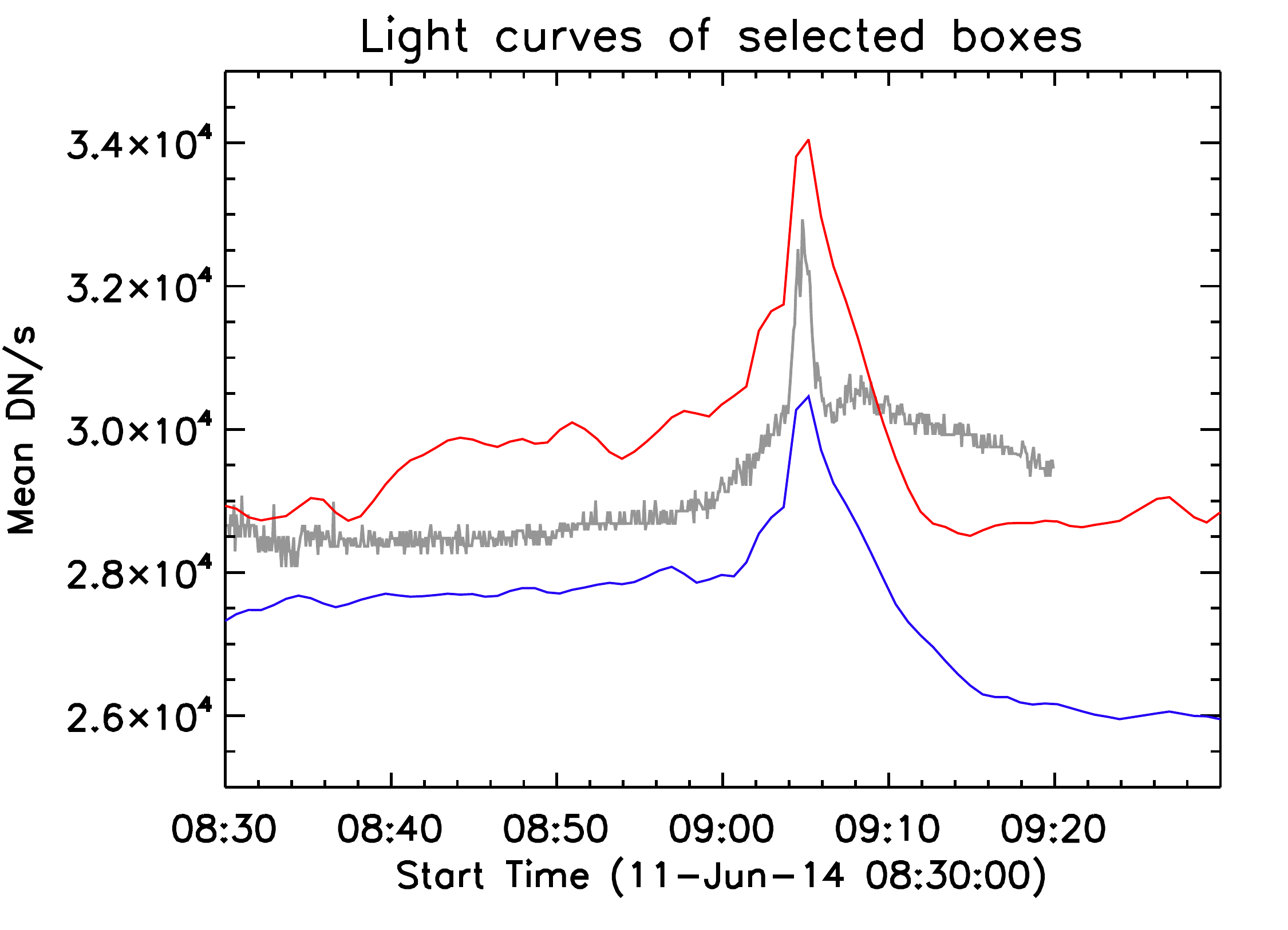}
\caption{Green contours highlight the white-light ribbons observed by SDO/HMI. A time evolution of the mean value over the red and blue boxes is plotted on the right panel. The orange contours mark the HXR sources with energies 40 - 70 keV observed by RHESSI at the time interval 9:04:14 - 9:04:40 UT. The grey lightcurve shows the corrected count rate of the RHESSI 50 - 100 keV channel in arbitrary units.}
\label{fig80}
\end{figure}

\textit{Helioseismic and Magnetic Imager} (SDO/HMI, \citealt{Scherrer:2012aa}) observations in the far wings of the Fe I line (617.3 nm) show evidence for white-light emission during the flare under investigation. We localize the source of the white-light emission with difference imaging using SSW routines. \software{hmi\_prep.pro, index2map.pro, drot\_map.pro and diff\_map.pro (\citealt{Freeland1998})}.\\
Once the bright kernels are localized, we cut two 5 pixel x 5 pixel subimages centered on the kernels and calculate their mean signal during the event - see Figure \ref{fig80}. Two ribbon-like structures are produced during the impulsive phase matching the maximum seen in the hard X-rays with a decay time around 8 - 10 minutes. The ribbons light up along their length from north-east to south-west and are co-spatial and co-temporal with the RHESSI hard X-rays.\\

\subsection{SDO/EVE}

\begin{figure}[h]
\plottwo{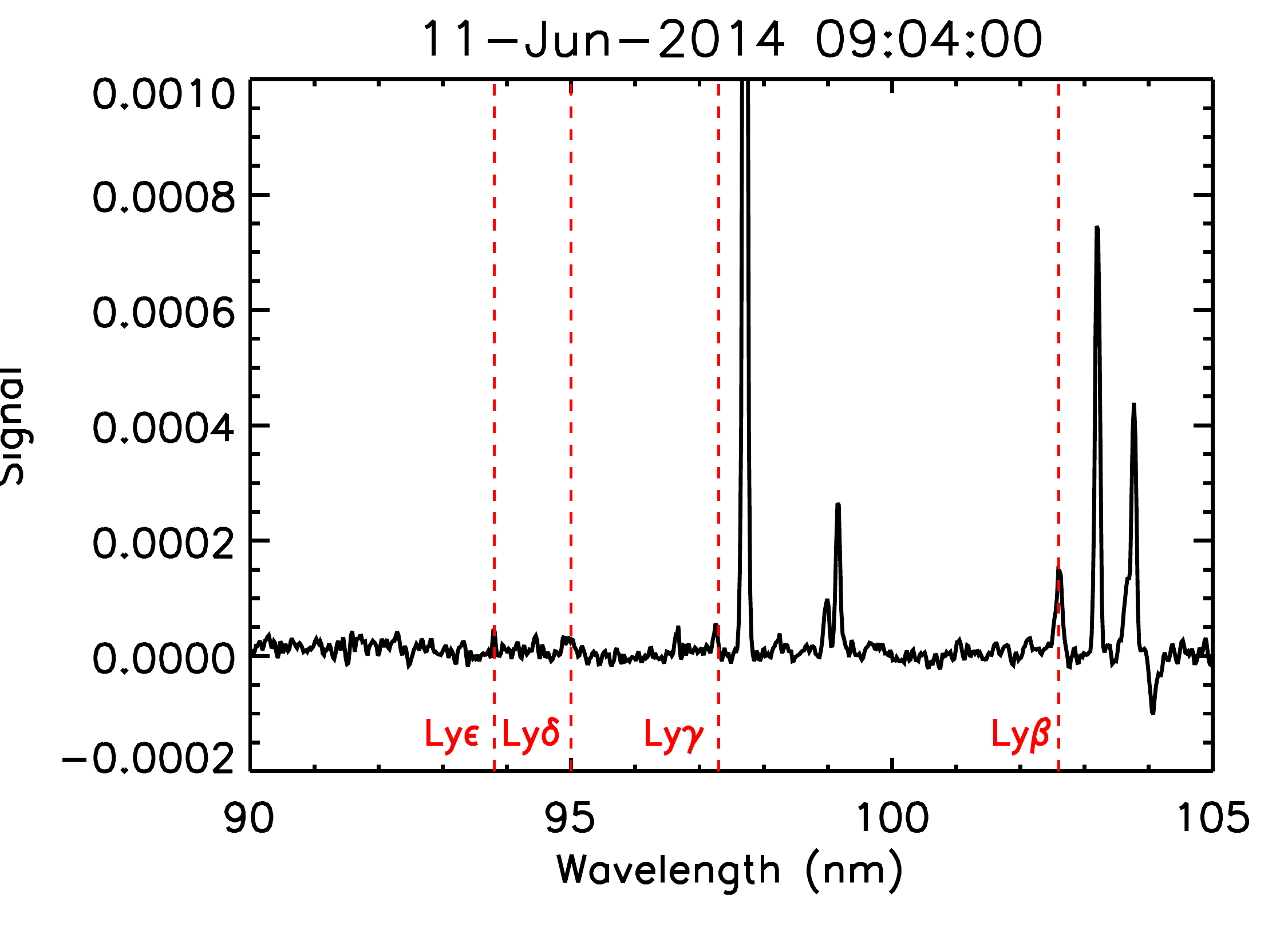}{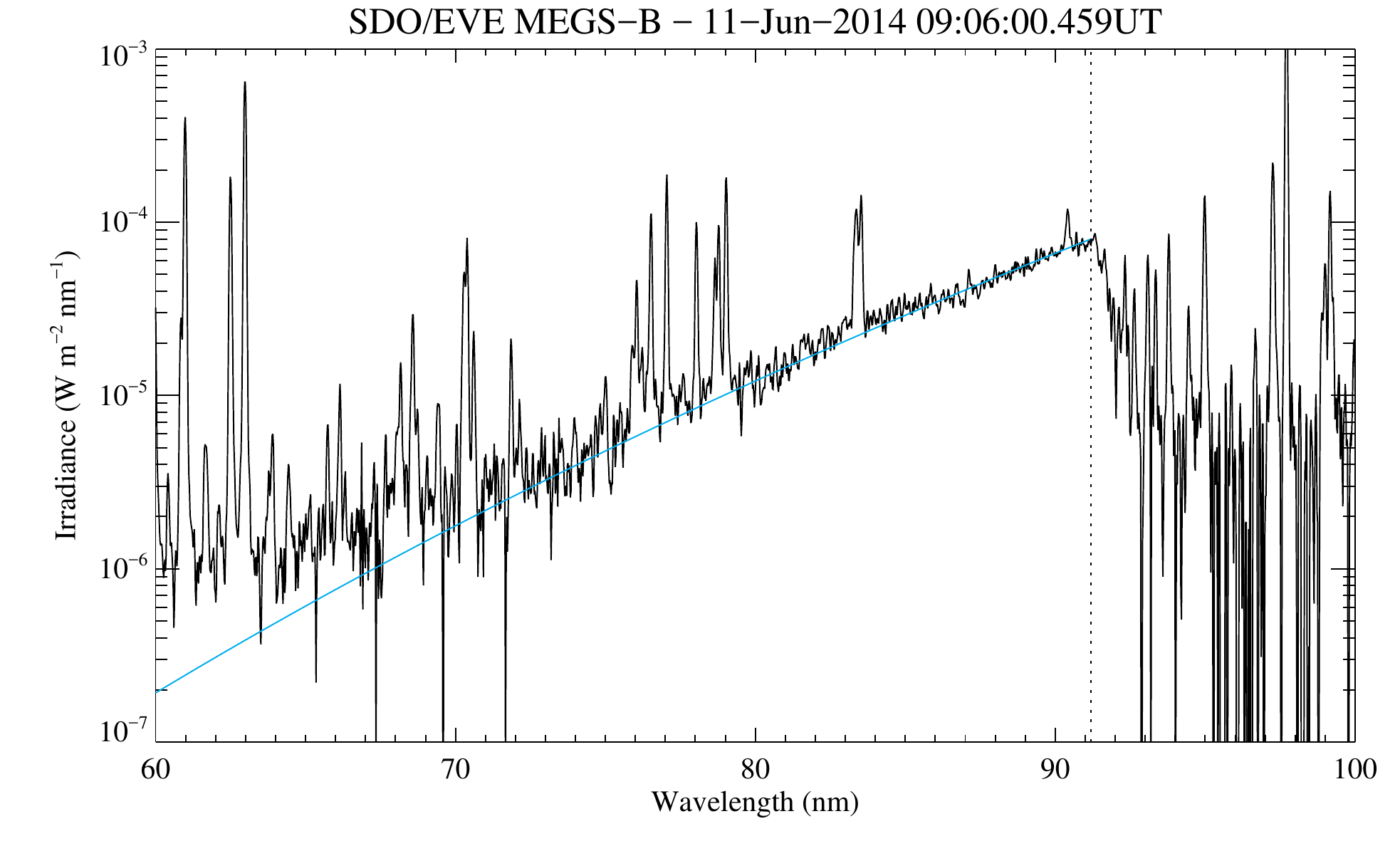}
\caption{Left: The flaring difference of the X1.0 event from SDO/EVE showing the strong suppression of the higher Lyman lines. Right: An example of SDO/EVE spectrum with a power-law fit (in blue) of Lyman continuum.}
\label{fig32}
\end{figure}

The \textit{EUV Variability Experiment} (EVE; \citealt{Woods:2012aa}) is one of three instruments onboard the \textit{Solar Dynamics Observatory} (SDO; \citealt{Pesnell:2012aa}). EVE measures the solar irradiance between 0.01 and 105.0 nm using \textit{Multiple EUV Grating Spectrographs} (MEGS) - A (5 - 37 nm), B (35 - 105 nm), Photometer (121.6 nm) and a EUV Spectrophotometer (0.01 - 3.9 nm) at a cadence of 10 seconds. The MEGS-B spectral range contains numerous emission lines formed over a broad range of temperatures. Many of the cooler lines - most notably Lyman $\beta$, Lyman $\gamma$, Lyman $\delta$ (hereafter Ly$\beta$, Ly$\gamma$, Ly$\delta$), are formed in the chromosphere, along with the Lyman free-bound continuum (LyC) with a recombination edge at 91.2 nm. A spectrum obtained during the impulsive phase of the X1.0 flare is shown in Figure \ref{fig32}. Due to unforeseen degradation, MEGS-B operates at a reduced duty cycle. Fortunately, on 11 June 2014 the instrument was exposed to the Sun for 3 hours between 08:00 and 11:00 UT, capturing both the M3.0 and X1.0 flares of that day.

In order to construct the temporal profile of the LyC for the events presented here, we applied the \textit{RANdom Sample Consensus} (RANSAC; \citealt{Fischler:1981}) technique described in \cite{Milligan14} to Version 5 of the EVE data. This technique treats any overlying emission lines as outliers in the data, allowing the `pure' continuum to be fit with a chosen function. The EVE data between 80 nm and the recombination edge at 91.2 nm were fit with a power law function that was extrapolated to shorter wavelengths (right panel of Figure \ref{fig32}). By integrating under the fit at each 10 second interval, lightcurves of LyC emission were established (top panel of Figure \ref{fig42}). Lightcurves of higher order Lyman lines were obtained by fitting each line with a Gaussian profile and integrating over the fit to get the total flux at each interval (middle panel of Figure \ref{fig42}).

While the MEGS-P diode provides measurements of the solar Ly$\alpha$ flux, also at 10 second cadence, \cite{Milligan:2016aa} showed that the temporal behavior of this emission during flares appears much more `gradual', rather than bursty as one would expect from an impulsively heated chromosphere. The reason for this unusual behavior is not clear. As such we shall use the Ly$\alpha$ lightcurves from the GOES/EUVS instrument instead (see Section \ref{EUVS}).

%

\subsection{GOES/EUVS Ly$\alpha$}
\label{EUVS}
\begin{figure}[p]
\centering
\vspace{-1.5cm}
\includegraphics[width=0.95 \textwidth]{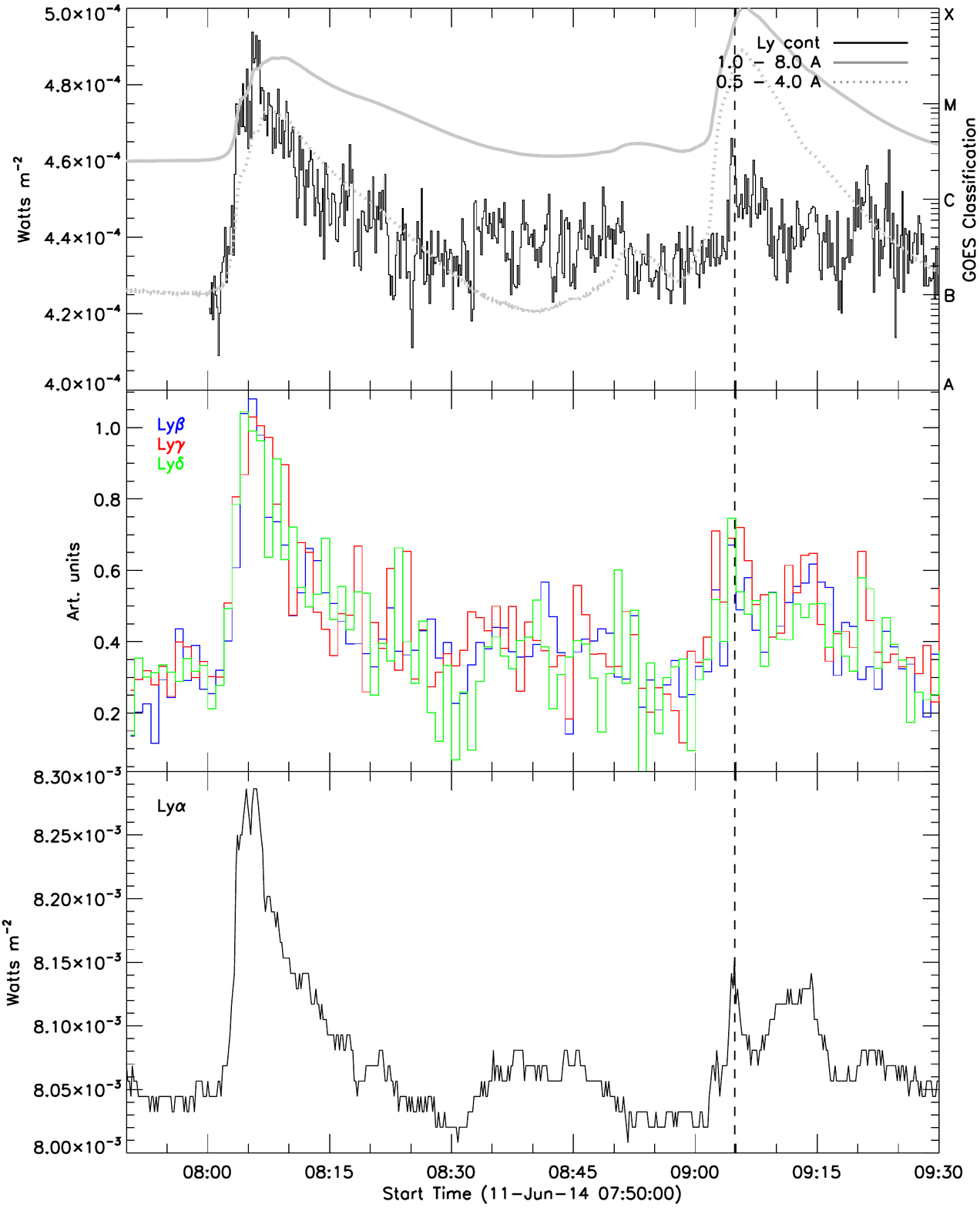}
\vspace{-1.5cm}
\caption{Light curves of Lyman continuum recorded by EVE/SDO and X-rays recorded by GOES (upper panel), Lyman lines (middle panel) recorded by EVE/SDO and Ly$\alpha$ (bottom panel) recorded by GOES. The vertical dashed line marks the maximum in RHESSI 100 - 300 keV flux.}
\label{fig42}
\end{figure}

As the Ly$\alpha$ data from SDO/EVE are somewhat inconsistent for reasons that are yet not fully understood, we utilized data from the EUV Sensor on GOES 15 instead. The EUVS onboard the previous three GOES satellites (13, 14, and 15) comprise five EUV channels: A, B, C, D, and, E. The E channel spans the Ly$\alpha$ line at 121.6 nm in a broadband ($\sim$10 nm) manner similar to MEGS-P, although its time profiles exhibit a behavior similar to that of LyC from EVE (\citealt{Milligan:2016aa}). The Ly$\alpha$ lightcurves for the M and X class flares on 11 June 2014 are plotted on the bottom panel of Figure \ref{fig42}. Similar to that seen in the EVE line and continuum data during the X1.0 flare, the Ly$\alpha$ emission was heavily suppressed relative to the M3.0 flare an hour earlier.

\section{Results}
\begin{figure}[h]
\vspace{-1cm}
\includegraphics[width=0.6 \textwidth]{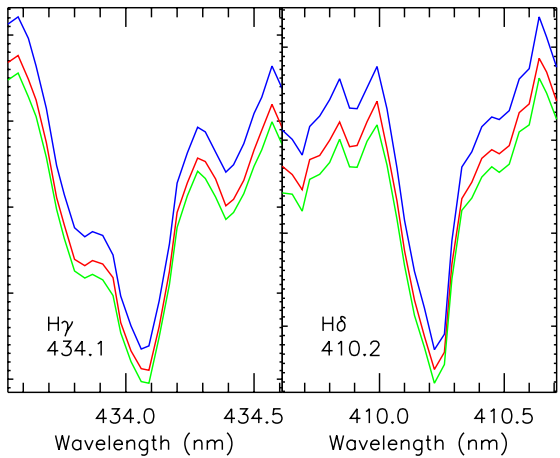}
\caption{Spectral profiles of H$\gamma$ and H$\delta$ during the rise (red), impulsive (green) and decay (blue) phase of the X1.0 flare on 11 June 2014.}
\label{fig83}
\end{figure}

The Ly$\alpha$ emission begins to rise together with hard X-rays at approximately 09:01 UT (Figure \ref{fig42}). Figure \ref{fig32} shows flaring spectra (only excess signal from the flare) of the X1.0 event where the higher hydrogen Lyman lines are heavily suppressed. A Lyman jump is not detectable in these observations (also see upper panel of Figure \ref{fig42}). The same figure shows a much stronger signal in the Lyman lines during the weaker M3.0 event. The flares of 11 June 2014 have a very weak response in the higher Balmer lines as these maintain their absorption profiles throughout the events. This effect is more pronounced in the X1.0 which shows no response at all (Figure \ref{fig83}). The same event shows brightening in the continuum at wavelengths $<$ 400 nm at 8:48 UT. The continuum brightening appears only about 1 - 2 minutes before the $<$ 25 keV burst was detected by RHESSI and remains elevated until the end of the observation at 9:10 UT, whereby its intensity varies between 10 - 21\% above the quiet level (see lower panel of Figure \ref{fig2}). Weak line core emission is detected in the Ca II H\&K lines.\\

Clear ribbon-like WL emission is visible on the HMI images but due to the low cadence (45 s) it is difficult to determine its exact timing. However, the observations show a good correspondence between the brightest WL emission and the hard X-rays ($>$ 40 keV). Table \ref{tab} shows a summary of the main observational characteristics of the four flare events observed on active region 12087 on June 10-11 2014. Although all X-class events reached similar flux in X-rays, the morphology of WL emission and line profiles are qualitatively different.

\begin{table}
\centering
\begin{tabular}{cccccc}
Peak time (UT) & Position & GOES classification & Balmer line profiles & WL emission & C364 nm  \\
\tableline
10 June 2014 & && & &\\
11:42 & S19E81 & X2.2 & Emission, CR pronounced & Very strong & $>$ 16\%\\
12:52 & S20E89 & X1.5 & Emission, minor CR & Widespread & $>$ 13\%\\
\tableline
11 June 2014 & && & &\\
08:09 & S18E68 & M3.0 & Absorption & Weak, points-like & None\\
09:06 & S18E66 & X1.0 & Absorption & Strong & 19\%\\
\end{tabular}
\caption{A summary of the main observational properties of the four white-light flares observed in NOAA 12087 on 10-11 June 2014. The C364 nm column shows the excess signal in the vicinity of the Balmer edge relatively to the non-flaring signal. For the first two flares we determine a lower limit due to a lack of measurements during impulsive phases. CR denotes the presence of a central reversal in the line profiles.}
\label{tab}
\end{table}

\section{RADYN Modelling}
\label{RADYN}
In order to understand the suppressed hydrogen Balmer and Lyman emission observed in the X1 flare, we carried out radiative hydrodynamic simulations. \software{RADYN (\citealt{Carlsson:1992aa,Carlsson:1995aa,Carlsson:1997aa} and \citealt{Allred:2015aa})}.\\
 The code has been used extensively to model the emission of both solar and stellar flares (e.g., \citealt{Abbett:1999ab}, \citealt{Allred:2005ab}, \citealt{Testa:2014aa}, \citealt{Kennedy:2015aa}, \citealt{Rubio-da-Costa:2015aa}, \citealt{Rubio-da-Costa:2016aa}, \citealt{Kuridze:2015aa}, \citealt{Kowalski:2016aa}, \citealt{Kerr:2016aa}). RADYN solves the coupled set of equations describing hydrodynamics, radiative transfer and non-LTE atomic level populations for a six-level with continuum H atom, nine-level with continua He I and He II ions,  and a six-level with continuum Ca II ion. It uses an adaptive grid (\citealt{Dorfi:1987aa}) so that it can resolve the high-speed shocks that develop during flares.  It models one-dimensional loop structures that extend from the sub-photosphere into the corona. To model the response of the solar atmosphere to flares, it has been coupled to an additional code that simulates the kinetic transport of non-thermal (i.e., flare-accelerated) particles injected at the loop top as described in \cite{Allred:2015aa}. This code models the accelerated particle distribution function in response to energy loss and pitch-angle scattering due to Coulomb collisions, synchrotron emission and magnetic mirroring. It includes relativistic effects, which are important for high energy particles. Energy lost by the non-thermal particles is assumed to be transferred to the ambient plasma in the form of heat. Additionally, when these particles collide with ambient ions, they may further ionize the ions. We have included that effect in the collisional ionization rate equations solved by RADYN using the methods of \cite{Fang:1993aa} and \cite{Arnaud:1985aa}.  Importantly for this work, RADYN can be used to predict the Balmer emission originating from optically-thick regions of the atmosphere in response to flare heating. Since RADYN uses a six-level hydrogen atom it directly predicts Balmer line profiles up to H$\gamma$. To model higher order Balmer lines, we have input snapshots of the loop temperature, density, and velocity as a function of column mass into the radiative transfer code, RH (\citealt{Uitenbroek:2001aa}). RH has been configured to include a 20-level model hydrogen atom allowing the prediction of line profiles up to H18.

\begin{figure}[h]
\centering
  \includegraphics[width=0.7 \textwidth]{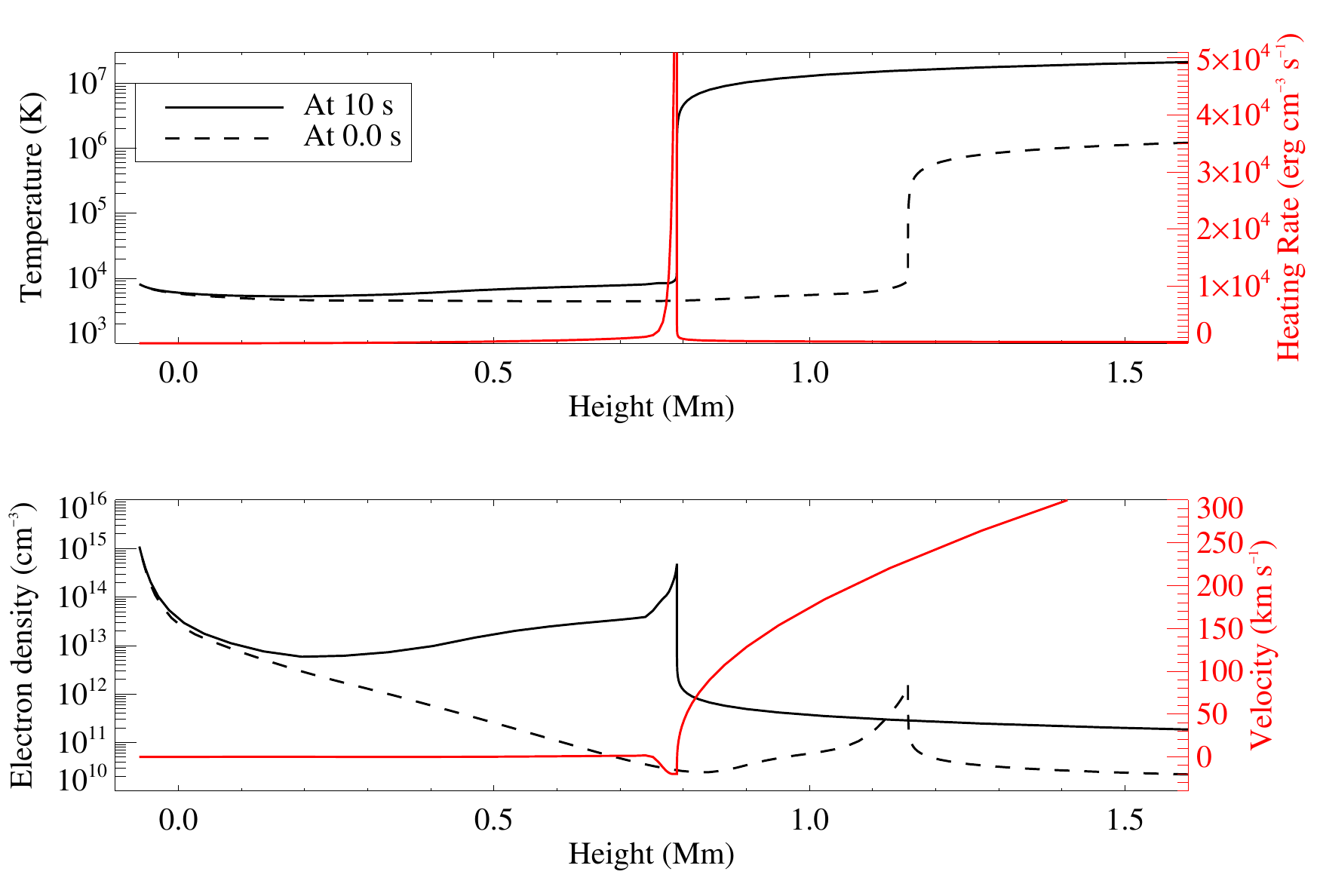} 
\caption{(Top panel) Temperature (black solid) and heating rate (red) as a function of height above the photosphere after 10 s of electron beam heating compared with the temperature in the initial loop (black dashed). (Bottom panel) Electron density (black solid) and velocity (red) as a function of height above the photosphere after 10 s of heating compared with the density in the initial loop (black dashed).}
\label{figJ1}
\end{figure}

We have configured RADYN to perform two experiments. In each case, heating was applied to a loop structure, which started in a state of hydrostatic equilibrium. We chose as a starting state the QS.SL.HT loop described in \cite{Allred:2015aa}. This loop has a half-length of 10 Mm, a constant cross-sectional area, and apex temperature and density of 3 MK and 10$^{10}$ cm$^{-3}$, respectively. In order to model the effects of magnetic mirroring and synchrotron emission on non-thermal particles, we have assumed a magnetic field strength which decreases exponentially from 1000 G in the footpoint to 100 G at the loop top. For the first experiment, we have injected electrons at the loop top. RHESSI hard X-ray observations taken during this flare were used to constrain the accelerated electron spectrum. The spectrum was best fit to a power-law with a cutoff energy, $E_c$, of 20 keV and a power-law index, $\delta$ = 3. The pitch angle distribution of the injected electrons was chosen to be a Gaussian centered around the loop axis and with a half width half max of 23.5 degrees. That $\delta$ is unusually low and indicates the presence of a relatively large number of high energy electrons. These penetrate deeply, so perhaps they could be responsible for heating the low chromosphere, below the region where Balmer emission originates. In this experiment, a flux of $3 \cdot 10^{11}$ erg cm$^{-2}$ s$^{-1}$ of electrons were injected continuously for 10 s. The atmosphere is quickly heated in response. The temperature and density along the axis of the loop are shown in Figure \ref{figJ1}. Additionally, we have plotted the heating rate due to the injected electron beam and the plasma velocity. The corona has been heated to 20 MK and has an electron density of 10$^{11}$ cm$^{-3}$ as a result of chromospheric evaporation which has brought material into the corona. The beam heating peaks at 0.79 Mm above the photosphere. In fact, the temperature has been increased nearly all the way to the photosphere. The Balmer emission predicted from this simulation is discussed in Section \ref{discussion}.

\begin{figure}[h]
\centering
  \includegraphics[width=0.7 \textwidth]{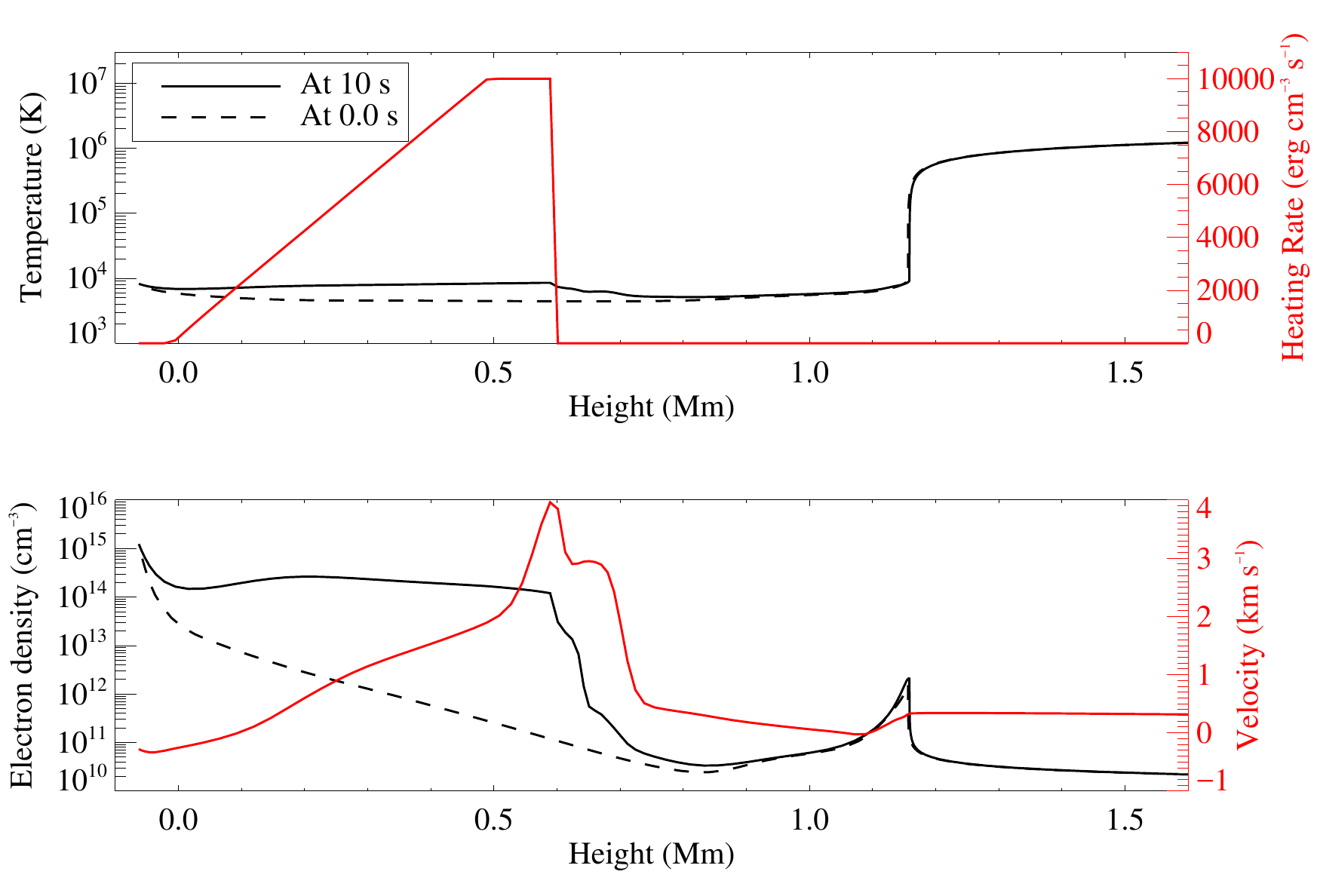} 
\caption{The quantities plotted are identical to those in Figure \ref{figJ1}, except that these are in response to direct temperature minimum heating. }
\label{figJ2}
\end{figure}

For the second experiment, rather than injecting electrons at the loop top, we applied heating directly to the temperature minimum region. A heating rate of $3\cdot10^{11}$ erg cm$^{-2}$ s$^{-1}$ was applied for 10 s, resulting in the same total energy deposited as in the first experiment. However, the dynamics of this simulation were quite different. The temperature, density, heating rate and velocity are plotted in Figure \ref{figJ2}.  In this case, since no electrons moved through the corona nor upper chromosphere, these regions remained relatively unaffected by the heating. However, the dense upper photosphere and temperature minimum region reach nearly 8000 K. The velocity has a maximum of only 4~km~s$^{-1}$.  The Balmer emission predicted from this simulation and compared with the previous experiment is discussed in Section \ref{discussion}.\\

\begin{figure}[h]
\includegraphics[width=0.5\textwidth]{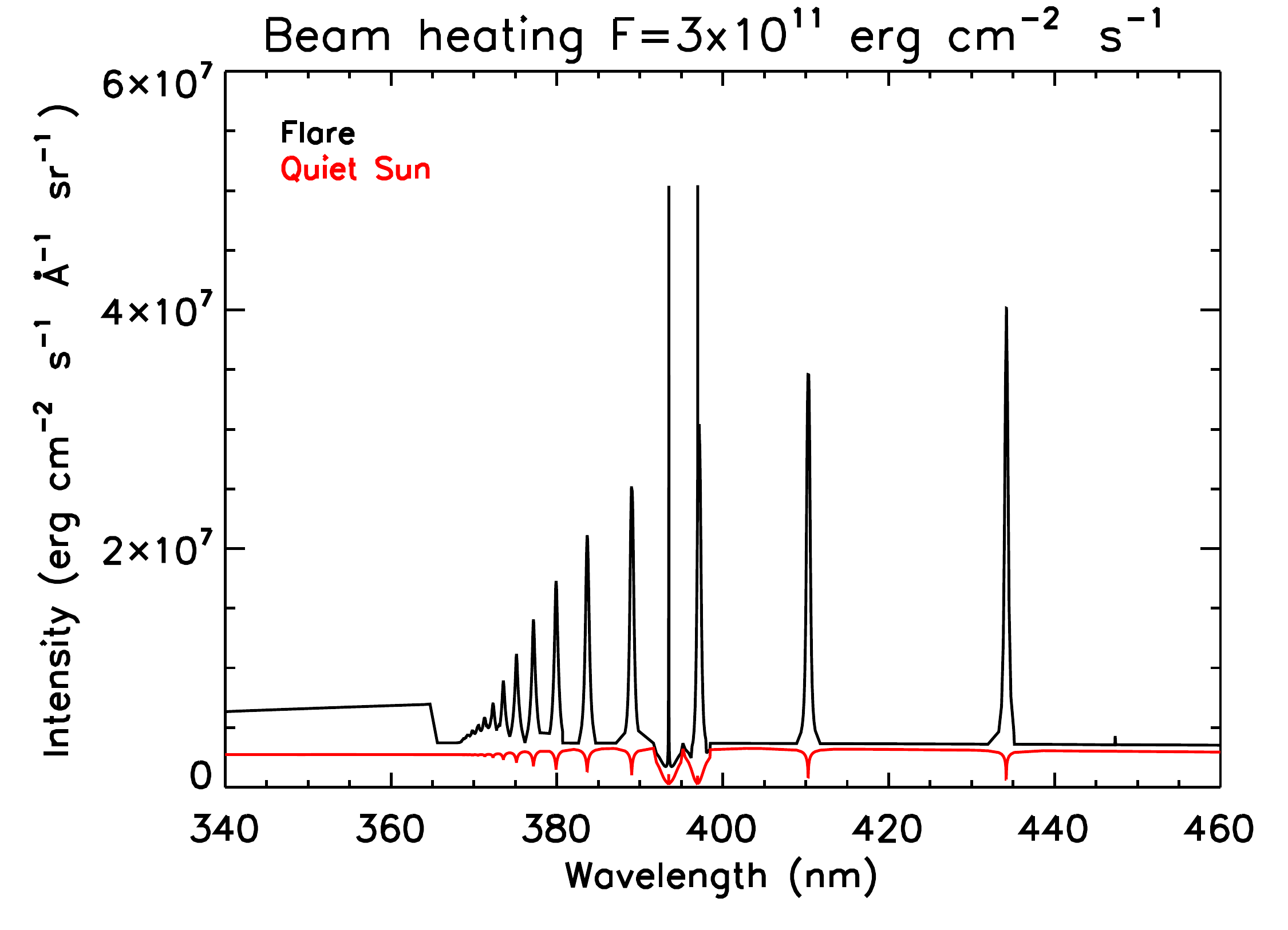}  
\includegraphics[width=0.5\textwidth]{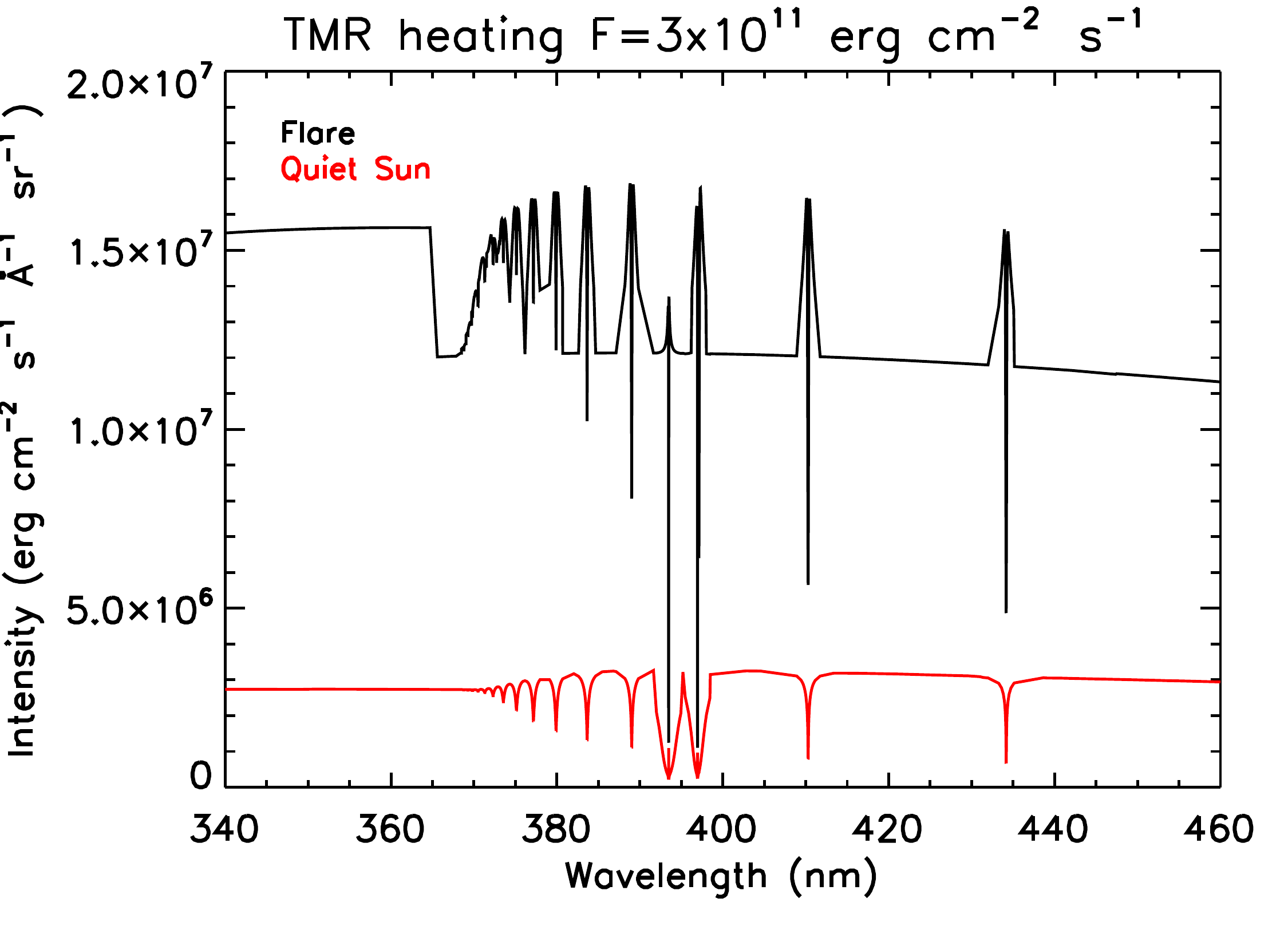} 
\includegraphics[width=0.5\textwidth]{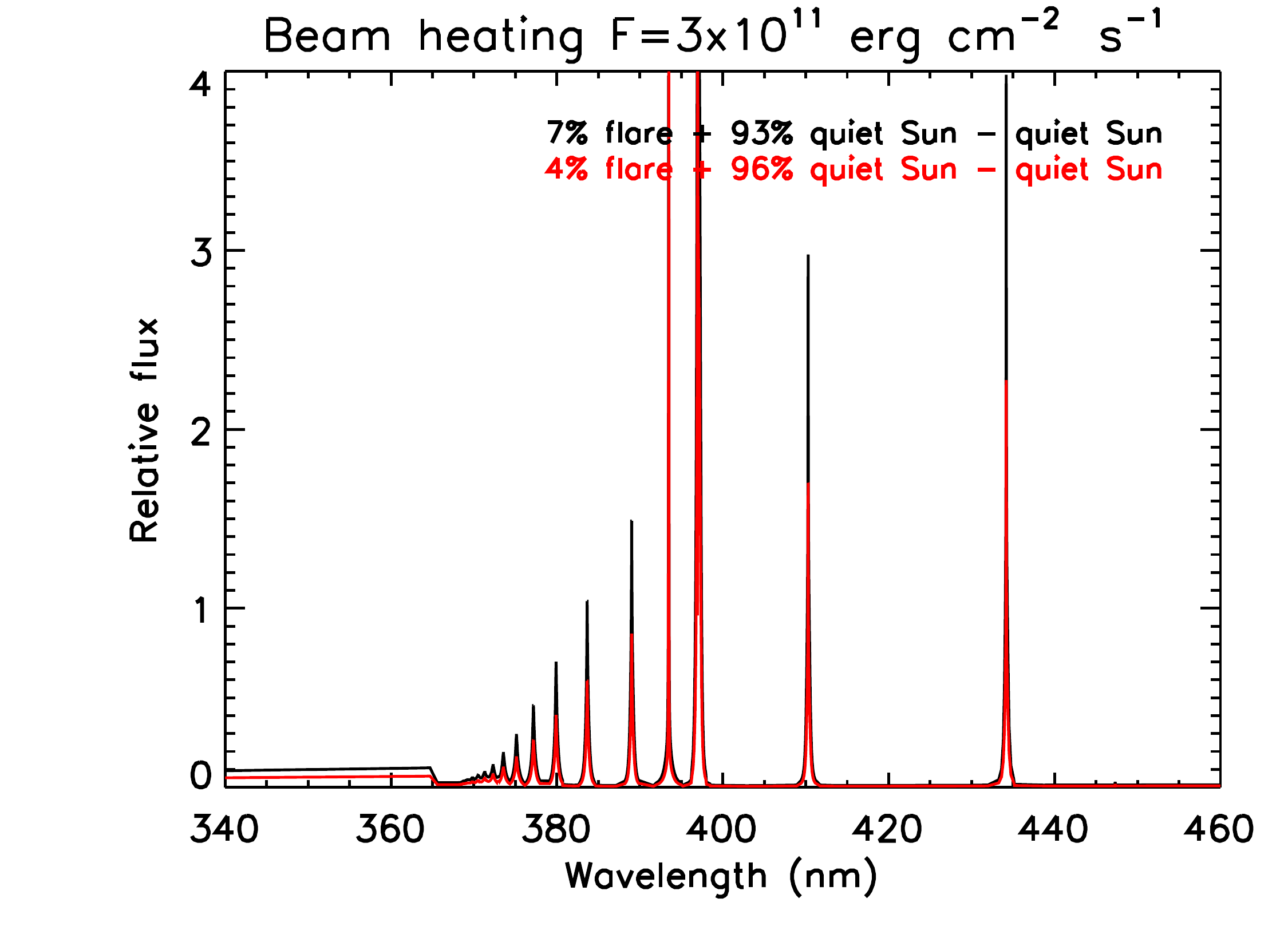}  
\includegraphics[width=0.5\textwidth]{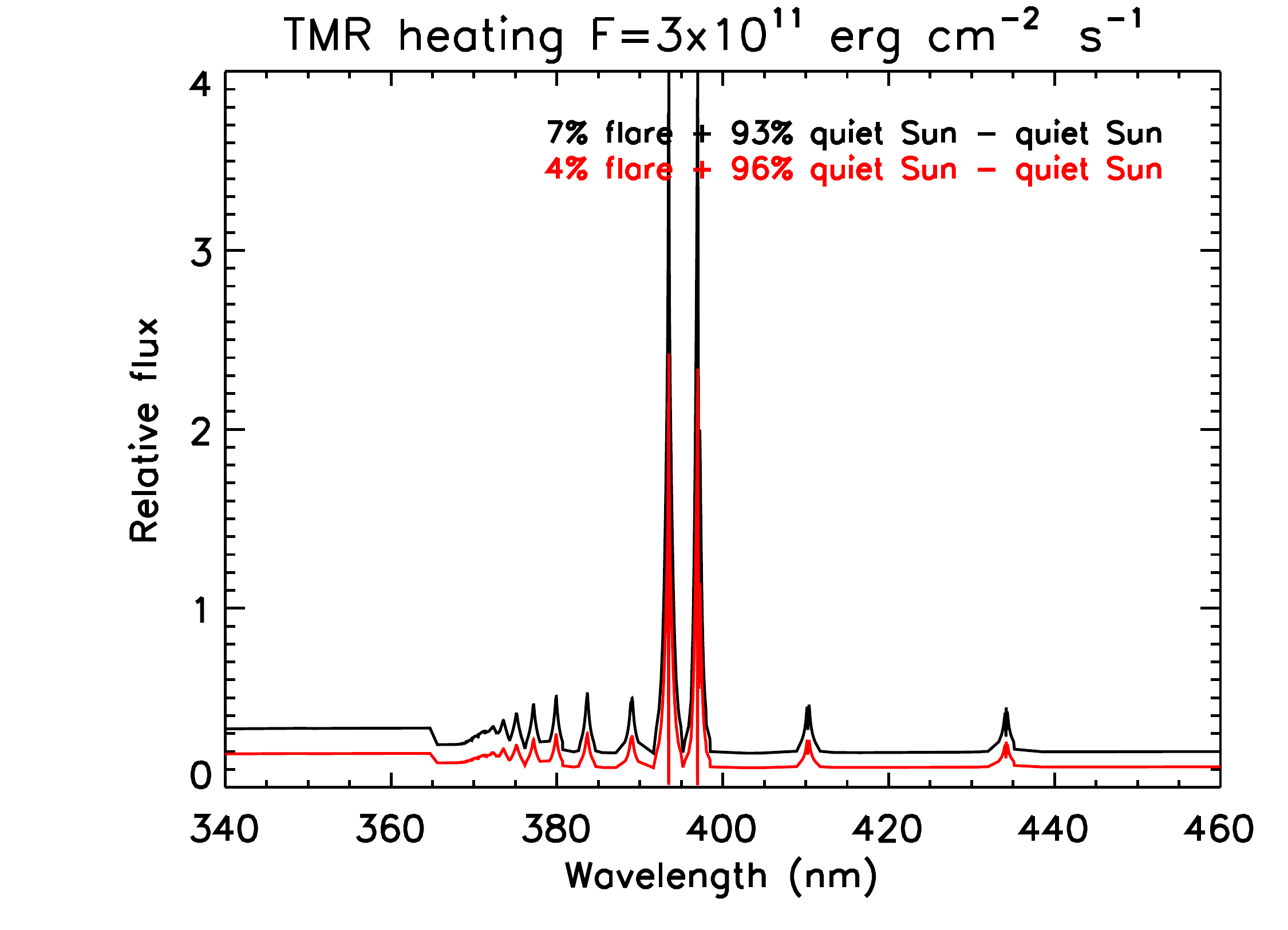} 
\caption{RH synthetic spectra produced from RADYN snapshots taken 10 s into the simulations. Upper left: A response of the atmosphere on electron beams ($F=3\cdot 10^{11}$ erg $\cdot$ cm$^{-2} \cdot$ s$^{-1}$, E$_c=20$ keV, $\delta=3$), upper right: A response of the atmosphere on direct temperature minimum region heating ($F=3\cdot 10^{11}$ erg $\cdot$ cm$^{-2} \cdot$ s$^{-1}$), lower left (right): a combination of flaring and quiet signal (see legend) divided by the quiet signal from beam (direct TMR) heating models.} 
\label{fig9}
\end{figure}

\section{Discussion \& Concluding Remarks}
\label{discussion}

\begin{figure}[h]
\centering
\includegraphics[width=0.8\textwidth]{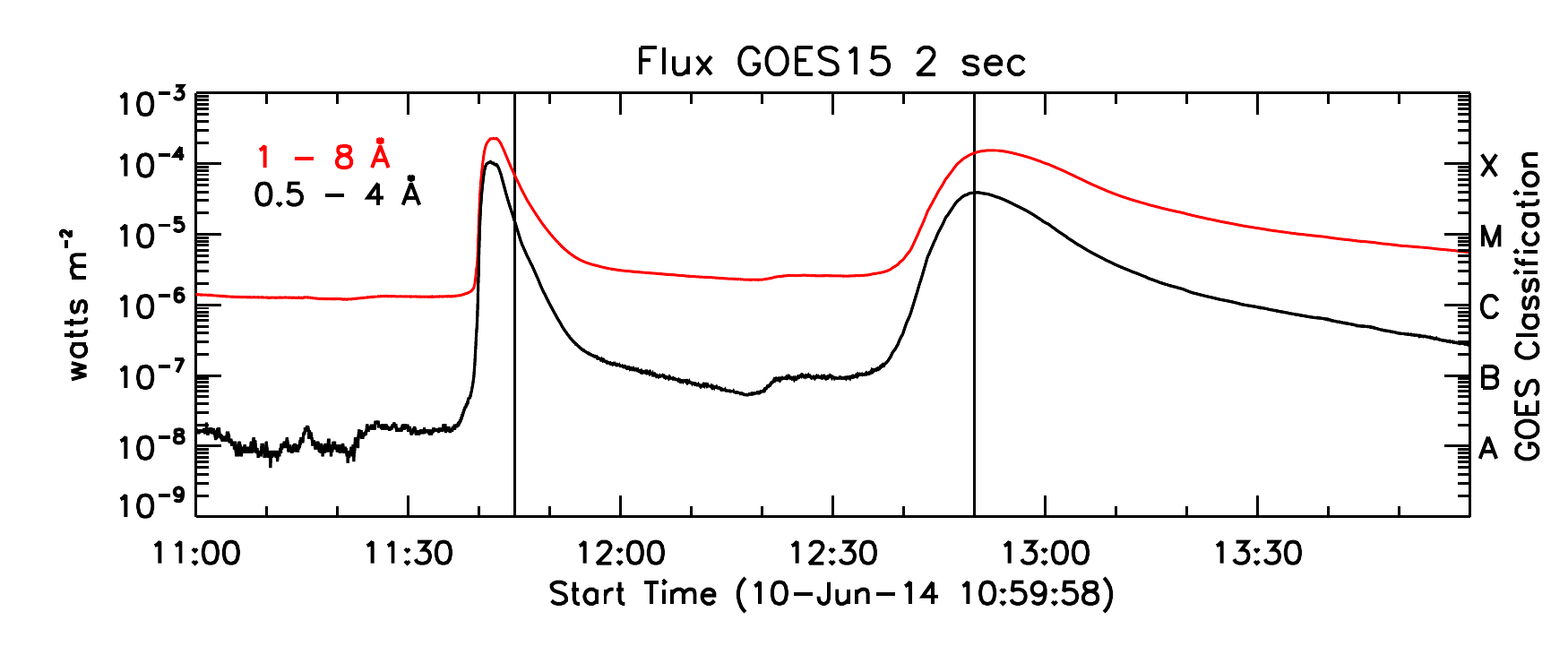}
\includegraphics[width=0.8\textwidth]{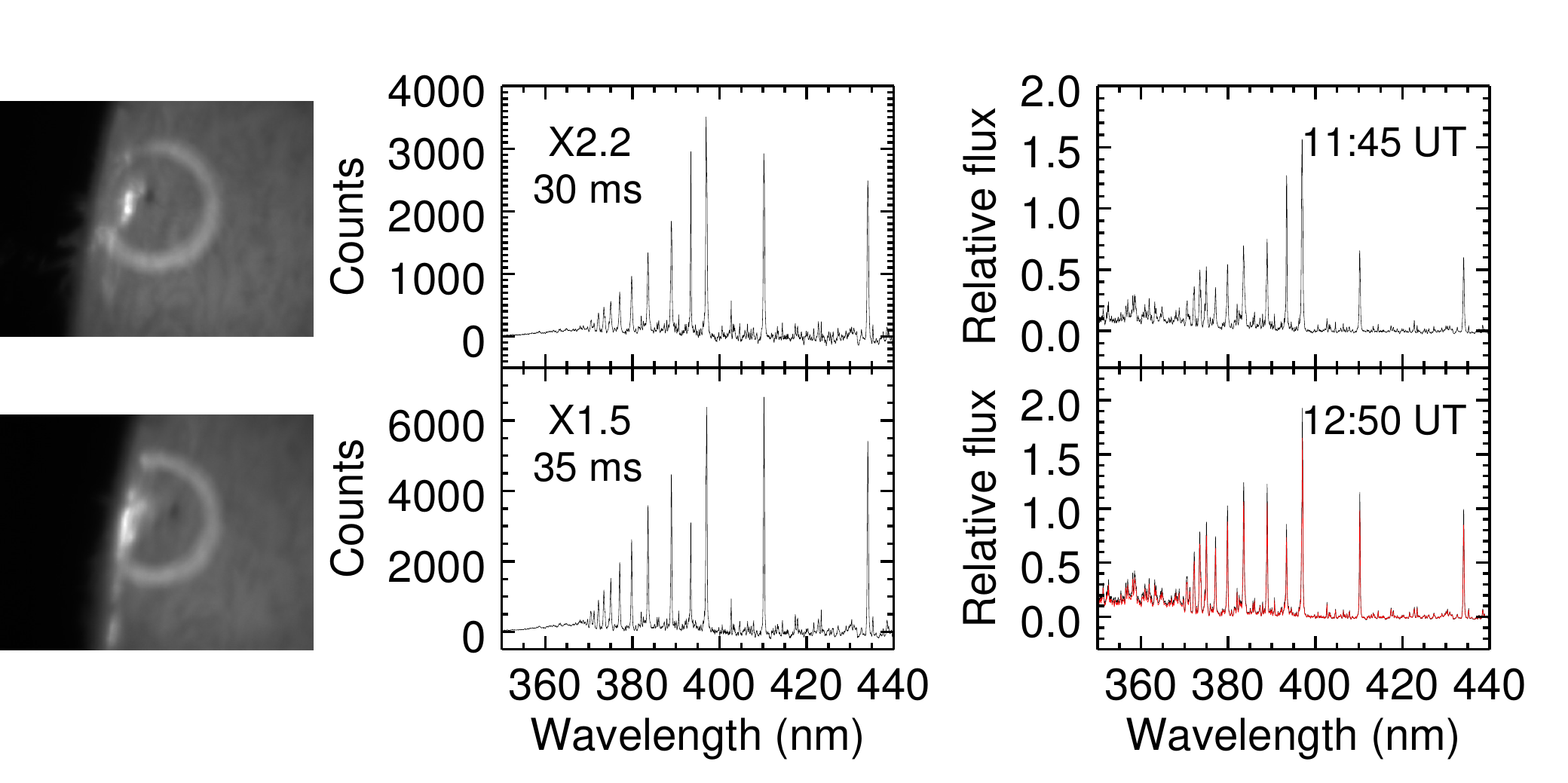}
\caption{Observations of X2.2 and X1.5 flares on 10 June 2014 with clear emissions in the Balmer lines (description as in Figure \ref{fig2}). The red spectrum was rescaled to match an exposure time 30 ms.}
\label{fig3}
\end{figure}
The context H$\alpha$ images in Figure \ref{fig2} show that the flare occurred in a relatively small part of the measured atmosphere inside the bright circle where the observed spectra have been integrated over. We estimated the flaring area by setting up an intensity threshold two times higher than the median signal in the image. We investigated a few images recorded during impulsive and early gradual phases. Due to seeing effect and evolution of the flare this yielded lower and upper estimates of 4\% and 7\%, for the flare area respectively. We synthesized the resulting spectrum for a 4\%, 7\% flare and 96\%, 93\%  quiescent Sun contributions respectively.  Figure \ref{fig9} shows that beam heating produces the well known strong emission in the Balmer line cores, while heating of the temperature minimum produces an increased continuum with weak core emission and enhanced line wings. The latter is qualitatively most similar to our observations (Figure \ref{fig2}). 
Our findings are in agreement with the limb flare observations of {\cite{Martinez-Oliveros:2012aa} who concluded that the WL emission and HXR emission originate at  a height of about 200 - 400 km which is deeper in the atmosphere than in our second modeling approach. Additional simulations show that depositing the energy into depths of 300 or 400 km produce clear absorption line profiles (Proch\'{a}zka et al. in preparation).   

Spectroscopic observations of solar flares in the blue part of the electromagnetic spectrum are rare. One such observation of C1.1 was recently presented by \cite{Kowalski:2015ab} who found emission in the higher order Balmer lines but no evidence for a Balmer jump. Their observations show a similar continuum shape in the vicinity of the Balmer jump to the one we present in the right column of Figure \ref{fig2}. \cite{Kleint:2016aa} used the Facility Infrared Spectropolarimeter (FIRS, \citealt{Jaeggli:2010aa}), Interface Region Imaging Spectrograph (IRIS, \citealt{De-Pontieu:2014aa}), HMI and RHESSI to investigate the energetics of an X1 flare.  Their modeling revealed the prevailing emission arised in the UV, visible and IR wavelengths. As blackbody emission alone does not fit the observed continuum, both blackbody and hydrogen recombination continua must be taken into consideration. In a unique stellar flare observation, \cite{Kowal2013} showed strong Balmer absorption lines (see their Section 6.3) during a megaflare on the M dwarf star YZ CMi. The spectral signatures of the megaflare resemble that of an A star spectrum which led to the conclusion that this is caused by a combination of a hot blackbody plus absorption component.

The spectra of the X-class flares observed from the same active region on 10 June 2014 show strong Balmer line emission (Figure \ref{fig3}). This is a common characteristic of solar flares \citep{Svestka:1972aa} and is in agreement with the findings of \cite{Kowalski:2015aa} who studied the observational signatures of both solar and dwarf M star flare atmospheres. Besides the strong emission lines, we also detected an excess continuum signal in shorter wavelengths ($<$ 410 nm) compared to the quiescent spectrum. For the X2.2 event we estimated the continuum excess as 16\% while for the X1.5 event it was 13\%. As we can see from the GOES X-rays, the X2.2 and X1.5 events were first captured during the early decay phase, so we accept these excess numbers as lower estimates.

Although an electron beam model can not be a priori excluded from the interpretation of our observations, the parameters of such a beam may have to be rather extreme if we were to consider electrons only. Lower energy electrons contribute to the heating of the upper chromosphere creating strong hydrogen line emission which we did not detect in the X1.0 event. Electrons with energies of 350 keV or above may be required to reach the temperature minimum region \citep{Aboudarham:1986aa}. The RHESSI data show evidence for particles of such a high energy in the studied flare, lower energy electrons are also present. Alternative mechanisms to transport the energy from the reconnection site to the lower layers will therefore also have to be considered. 

We believe that our observations combined with the radiative hydrodynamic simulations provide strong evidence that the heating of the X1 11 June 2014 flare occurs below the chromosphere. Given the lack or strong suppression of hydrogen Balmer  and Lyman lines, we speculate that this heating may not be a consequence of electron beams accelerated to the lower layers.
\cite{Zharkova:2007aa} compared standard electron beams and mixed proton/electrons beams to explain an X17.2 flare from 28 October 2003. They concluded that mixed beams produce shocks that help with the delivery of momentum into the deeper layers. An alternative mechanism presented by \cite{Fletcher:2008aa} proposes Alfv\'{e}n waves as a way for transporting the energy through the chromosphere. \cite{Russell:2013aa} claimed that the depth where the energy of the Alfv\'{e}n waves is deposited depends strongly on their frequency spectrum. Waves with periods longer than 10 s penetrate deeper the higher density parts of the lower atmosphere and can heat the temperature minimum region. 

The absence of emission in the higher order Balmer lines and the absence of a Balmer jump in the X1 class event together with the strong suppression of Lyman lines, are features of type II WLFs. Our radiative hydrodynamic models show that by the depositing the energy in the upper photosphere and temperature minimum region the spectral characteristics of these flares can be reproduced. 
\acknowledgments
The research leading to these results has received funding from the European Community's Seventh Framework Programme (FP7/2007-2013) under grant agreement no. 606862 (F-CHROMA). Ryan O. Milligan acknowledges support from NASA LWS/SDO Data Analysis grant NNX14AE07G. Pavel Kotr\v{c} acknowledges support from GA CR grant 16-18495S.
\bibliography{absence_of_lines_arXiv.bib}
\end{document}